\documentclass{article}

\usepackage[hidelinks]{hyperref}
\usepackage{breakurl}
\hypersetup{breaklinks=true}
\usepackage[preprint]{neurips_2023}
\usepackage[utf8]{inputenc}
\usepackage[T1]{fontenc}    % use 8-bit T1 fonts

\usepackage{booktabs}       % professional-quality tables
\usepackage{amsfonts}       % blackboard math symbols
\usepackage{nicefrac}       % compact symbols for 1/2, etc.
\usepackage{microtype}      % microtypography
\usepackage{amsmath,amsthm,enumitem} %for steps illustration
\usepackage{xspace}
\usepackage{graphicx}
\usepackage{subcaption}
\usepackage{todonotes}
\usepackage{xcolor}

\definecolor{darkblue}{HTML}{00008B}

\theoremstyle{definition}  % This style is typically for definitions and examples
\newtheorem{theorem}{Theorem}[section]  % Define a theorem environment

\theoremstyle{plain} 

\newcommand{\fullfledge}{\textit{Protected Audience}\xspace}
\newcommand{\fledge}{\textit{PrAu}\xspace}
\newcommand{\FOT}{\textit{FOT}\xspace}

\renewcommand{\vec}[1]{\boldsymbol{#1}}

\newcommand\shortsection[1]{\vspace{6pt}{\noindent\bf #1.}}

\newcommand{\keyword}[1]{{\sf #1}}

\title{Evaluating Google's~Protected Audience Protocol}

\author{
  Minjun Long \\
  University of Virginia \\
  \And
  David Evans \\
  University of Virginia\\
}
\date{\today}

\begin{document}

\maketitle

\begin{abstract}
While third-party cookies have been a key component of the digital marketing ecosystem for years, they allow users to be tracked across web sites in ways that raise serious privacy concerns. Google has proposed the \emph{Privacy Sandbox} initiative to enable ad targeting without third-party cookies. While there have been several studies focused on other aspects of this initiative, there has been little analysis to date as to how well the system achieves the intended goal of preventing request linking. This work focuses on analyzing linkage privacy risks for the reporting mechanisms proposed in the \fullfledge (\fledge) proposal (previously known as FLEDGE), which is intended to enable online remarketing without using third-party cookies.
We summarize the overall workflow of \fledge\ and highlight potential privacy risks associated with its proposed design, focusing on scenarios in which adversaries attempt to link requests to different sites to the same user. We show how a realistic adversary would be still able to use the privacy-protected reporting mechanisms to link user requests and conduct mass surveillance, even with correct implementations of all the currently proposed privacy mechanisms.
\end{abstract}

\section{Introduction} \label{sec:introduction} 

As designed, the HTTP protocol is stateless---each HTTP request from web browsers is treated independently \cite{fielding1999rfc2616}. To maintain state, web cookies were introduced. These allow a server to retrieve previously provided information, such as the contents of a user's shopping cart \cite{kristol2001http}. Cookies can be categorized as either \emph{first-party cookies}, placed by the website owner, or \emph{third-party cookies}, placed by a site other than the domain hosting the page the user visited. Third-party cookies can be used to track user activity across multiple sites. This enables ad targeting which increases revenue for publishers \cite{libert2018third, ravichandran2019effect}, but raises serious privacy concerns. To address the use of third-party cookies by companies, the European Union has adopted privacy regulations with strict penalties for non-compliance~\cite{mayer2012third} and the US Federal Trade Commission (FTC) has raised concerns~\cite{ftc2023} and taken action~\cite{ftc2012}.

%In response, 
Several popular web browsers have taken steps to safeguard user privacy by combating tracking. In 2018, Firefox revealed its plans to block third-party cookies based on tracking domains~\cite{firefox2018}. In 2020, Safari became the first widely-used browser to block third-party cookies by default \cite{safari2020}. Google also acknowledged the importance of protecting user privacy in a 2019 blog post~\cite{google2019}, where they improved cookie control in their Chrome browser. However, instead of immediately blocking all third-party cookies, they announced their intention to phase them out gradually through a plan they called the ``Privacy Sandbox''. This initiative aimed to find alternative ways to deliver personalized ads without compromising user privacy by phasing out the usage of third-party cookies, fingerprinting, and other advertising techniques that enable tracking web users across multiple sites \cite{sandbox2020}. While the original plan was to deprecate third-party cookies by early 2022, Google delayed this process to late 2023, partly due to regulatory pressure from the UK’s Competition and Markets Authority (CMA) \cite{phaseout2023}. Later, Google postponed the date again to the second half of 2024 \cite{phaseout2024}, as they have received constant feedback that more time is needed to evaluate and test the new Privacy Sandbox technologies before deprecating third-party cookies in Chrome. As of May 2024, Chrome still does not block third-party cookies by default.

Google's Privacy Sandbox initiative comprises several components including the Private State Token API for fighting fraud and spam on the web, the Attribution Reporting API for enabling digital ads measurement, and the Fenced Frames API to strengthen cross-site privacy boundaries~\cite{timeline}. Among all the proposals in Google's Privacy Sandbox, most of the work in the privacy research community has focused on the Federated Learning of Cohorts (FLoC) proposal \cite{floc2021dev}, which analyzes users' online activity within the browser and group a given user with other users who access similar content. After getting feedback from the FLoC's Origin Trial, Google replaced the FLoC proposal with the Topics proposal \cite{topic2022} in January 2022, where the browser observes and records topics that appear to be of interest to the user based on their browsing activity. Ad platforms can then access a user's interests through an API without obtaining their detailed browsing history. 

In this paper, we focus on the \fullfledge\ (\fledge) component (originally called \emph{FLEDGE}, an acronym of \emph{First Locally-Executed Decision over Groups Experiment}, and renamed as \fullfledge\ in April 2023~\cite{namechange}). Since March 2022, Chrome has been running the First Origin Trial (\FOT) on \fledge, which enabled this feature for a subset of Chrome users. Sites can also request trial tokens to participate in the \FOT and thus experiment with the API~\cite{origintrial}. 

The \fledge\ proposal~\cite{fledge2022} redesigns the advertising ecosystem by letting the browser client, rather than the centralized market operator, maintain the information about the user's interests for ad targeting. It moves ad auctions from external servers to the client browser, thus avoids the need to send information that can be used to track a user to a centralized ad auction server. \fledge is intended to enable advertisers to target ads based on user interests, but without revealing information about the user---in particular, \fledge is designed to prevent user tracking by ensuring that advertisers and publishers cannot link requests to an individual, while enabling some degree of interest-based behavior targeting.

Although \fledge\ shares the goal of grouping users by interests with the Topics API, its approach and implications for privacy are markedly different. The Topics API facilitates a better understanding of the content that users are interested in without sharing specific user data, but recent studies have shown that it is it possible for adversaries, through statistical modeling and collaboration with third parties, to re-identify users across different websites \cite{jha2023robustness, beugin2023interest}. In contrast, \fledge is designed to enable advertisers to target specific user groups without disclosing individual user data but its privacy mechanisms and potential vulnerabilities have not been as thoroughly explored. Since there are several parties involved in the \fledge protocol, any part of the lifecycle may be compromised and result in weaker privacy protection than intended. Thus, it is worth examining the whole workflow of \fledge to evaluate how well it satisfies its key privacy goal of preventing request linking. We focus on the ad reporting mechanism in \fledge as this is the only explicit communication channel that sends user data back to centralized servers.
%, under the assumption that other parts of the workflow remain uncompromised. 

Although Google's most recent announcement set the date for removing third-party cookies in 2024 \cite{phaseout2024}, the currently available implementation of \fledge is only partial. Many components are not yet implemented including the Trusted Key/Value Service, the integration of \emph{k}-anonymity, and Trusted Aggregation Service, and much of the documentation is still incomplete. Our analysis and evaluation are based on the official \fledge API developer guide \cite{fledge2022} and experiments on the available \FOT implementation of the \fledge API.  In this sense, our security analysis is premature---we analyze a system that does not yet fully exist and for some aspects we must speculate on what the eventual design will be. Nevertheless, there is an urgent need for the privacy research community to independently and objectively evaluate privacy risks of the proposed \fledge design before it is deployed at a wide scale, and hope the results of such an evaluation will influence the eventual deployment of \fledge.

\shortsection{Contributions}
We summarize the \fledge protocol, describing the interactions through the \fledge\ APIs among its components (\autoref{sec:background}).  
To analyze how well the \fledge\ design meets its privacy goals, we introduce a threat model (\autoref{sec:threat}) and evaluate three scenarios in which advertisers may use \fledge\ auctions to track users between sessions (\autoref{sec:analysis}). We find that the aggregate reporting mechanisms provided by \fledge can be used to link requests with high accuracy, scaling up to a level where mass surveillance is possible (\autoref{ss:large-pool-target}) even when the proposed $k$-anonymity mechanisms are implemented (\autoref{sec:circumvent-k-anon}). Although there is no simple fix that provides both the desired reporting and unlinkability, we discuss several potential mitigations in \autoref{sec:countermeasures}.

\section{The \fullfledge Protocol} \label{ss:protocol}\label{sec:background}

The \fledge\ protocol enables on-device auctions that run in the user's browser instead of running on a remote server to support ad targeting (including remarketing, which lets advertisers customise their display ads campaign for users who have previously visited the website~\cite{remarketinginsta}) without third-party cookies. Potential ad buyers can perform on-device bidding based on interest group metadata and data loaded from a trusted server at the time of the on-device auction, and the sellers who own ad display space on the visited page can perform on-device ad selection based on bids and metadata entered into the auction by the buyers. By storing interest groups in the browser and moving the auction process to isolated browser worklets, one running code provided by the prospective ad buyer and one running code provided by the seller, \fledge aims to enable ad targeting without the privacy compromises associated with third-party cookies. The individual's interests are stored in their own browser and can be used for ad targeting without even needing to be revealed to the ad seller or a centralized auction server; the buyer can buy an ad to be displayed to the individual on the seller's site without learning about the site visited. After the auction ends, the winning ad will be rendered in a \emph{fenced frame}. Unlike iframes, fenced frames allow access to cross-site data \emph{without} sharing it with the embedding context~\cite{fencedframe}. Code running in a fenced frame cannot communicate with the embedding context and vice-versa, though it may leak information through side channels~\cite{fencedframesidechannel}.

\subsection{Protocol Overview} \label{sec:protocol-overview}

\begin{figure*}[ht]
    \centering
    \includegraphics[width=0.9\textwidth]
    {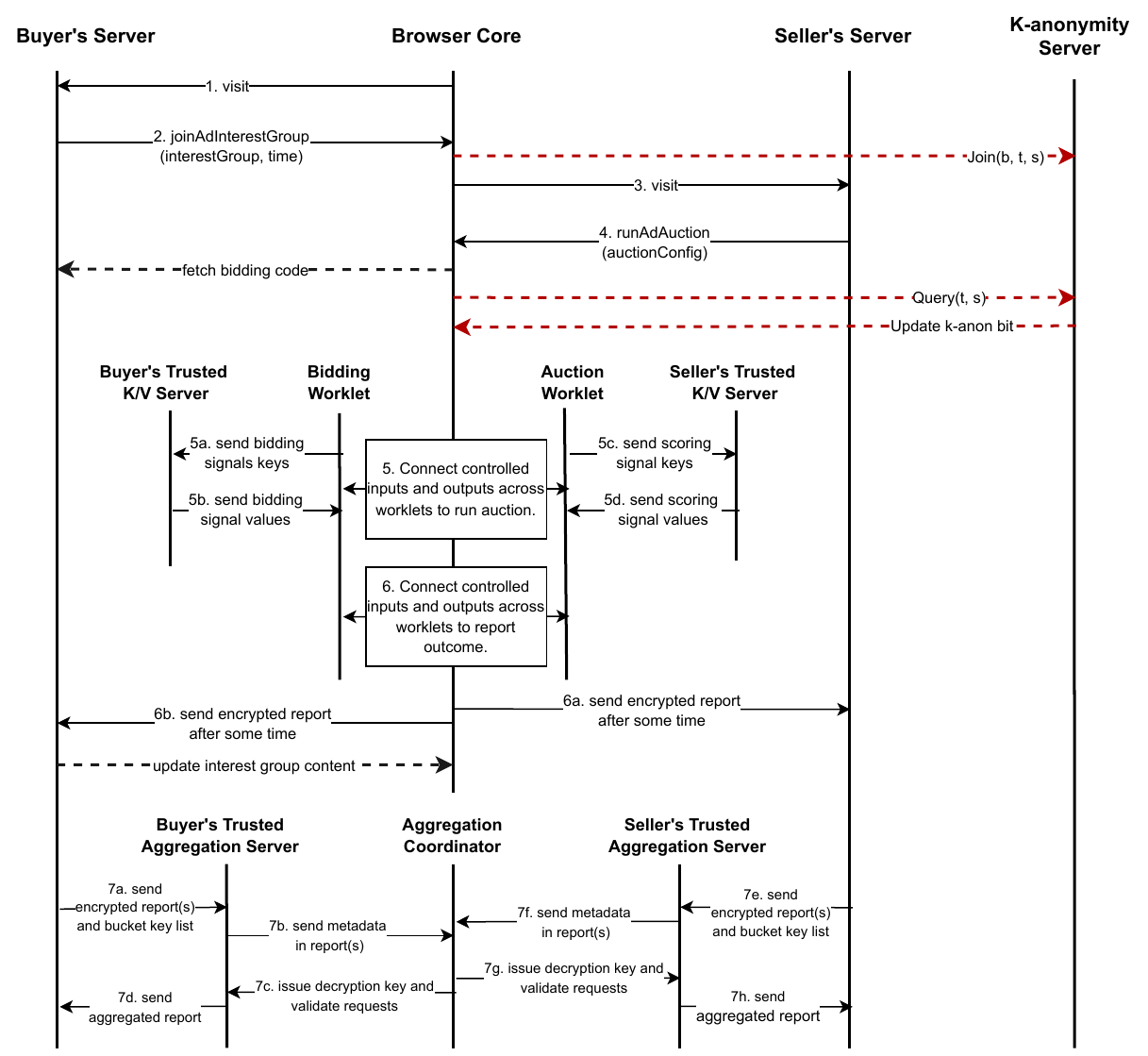}
    \caption{\fullfledge Protocol. \rm For simplicity, we show a single seller and buyer (who is also the winner of the auction), although there would typically be many buyers and sellers. Dashed lines indicate requests that do not necessarily happen at a particular step in the protocol, and may be interleaved with other request in different ways. The black dashed line for ``fetch bidding code'' is discussed at the end of \autoref{ss:protocol}. The black dashed line for ``update interest group content'' is requested daily, detailed in \autoref{ig}. The requests to the K-anonymity server are done periodically, details in \autoref{ssec:k-anon}. The controlled inputs and outputs across worklets are specified in \autoref{fig:auction-process}.} 
    \label{fig:design}
\end{figure*}

\autoref{fig:design} shows the design of \fledge. Our understanding of the protocol is based on publicly-available information, including Google's official documentation~\cite{fledge2022, fledgedev}, examining the code provided in associated github repositories~\cite{fledgegithub}, observing network traffic from the \FOT, answers to our questions from the Google team, and our own inferences regarding the planned design based on our best interpretation of the available material. \autoref{fig:FOT} (in the Appendix) shows what is currently implemented in the \FOT, which does not incorporate the Trusted Aggregation Server and other essential aspects of the planned design. Since Chrome is still supporting third-party cookies, the lack of privacy mechanisms in the currently deployed \fledge\ is reasonable since the tracking they are designed to thwart is fully enabled by the supported third-party cookies. However, once Chrome phases out third-party cookies, the adoption of \fledge's full suite of privacy mechanisms is intended to inhibit the type of request linking that third-party cookies currently enable. 

\fledge allows websites to store information about a user's interests in the form of interest groups that are stored in the browser and tied to the user profile for up to 30 days. Note that interest groups are tied to a browser \emph{profile} instead of a Google account, so are not shared across different browsers and devices even when they are connected to the same Google account. While the content of the group can be updated by the owner of the group by providing the name of the interest group, it cannot be read directly---interest groups are stored in an internal data structure and no API is provided to obtain a user's interest groups. When a user visits a site that sells ad space, buyers who are the owners of interest groups stored in the user's browser can participate in an auction. The in-browser auction will obtain some information from the buyer's code such as bid value and ad information, a score from the seller's code for each potential ad, and select the ad with the highest score to display.

The \fledge design relies on several \keyword{Trusted Servers} operated on the behalf of buyers and sellers. These services are trusted based on code reviews conducted by a trusted auditor and the servers that receive and respond to requests are required to run within a Trusted Execution Environment (TEE). A TEE is an isolated execution context that provides hardware memory protection and storage protection through cryptography, ensuring that its contents are shielded from observation and tampering by unauthorized parties, including the root user \cite{sabt2015trusted}. 

The interactions among buyers, sellers, and browsers in \fledge can be divided into the following main steps, depicted in \autoref{fig:design}:

\begin{enumerate}[leftmargin=4em,itemsep=1.5ex,labelwidth=1cm,labelindent=0.5cm,label=\bfseries Step \arabic*.]%,align=left]
\setcounter{enumi}{0}
\item The user visits a buyer site.
\item The buyer's script running in the browser calls \keyword{joinAdInter\-estGroup($\cdot$)}, which attaches an interest group to the user. 

\item The user visits the seller's site. 

\item The seller's script running in the browser calls \keyword{runAdAuc\-tion($\cdot$)} to initiate an auction (\autoref{auction-config}).

\item The browser runs auctions with each potential buyer's code in a separate bidding worklet and the seller's code running in the auction worklet (\autoref{sec:runningauction}). Code in worklets can send requests to Trusted K/V Servers to exchange real-time information \cite{kvservice}.

% Step 6
\item After a random delay of up to one hour, encrypted event-level reports on the auction results are sent to servers owned by the buyer and seller. 
% Step 7
\item Buyers and sellers submit encrypted event-level reports with queries to their respective Aggregation Server, which interacts with the Aggregation Coordinator to enable decrypting the event-level reports to produce and return a privacy-noised aggregated report (\autoref{aggregate-service}).

\end{enumerate}

Note that none of these privacy protections are implemented in the \FOT, which just immediately sends a cleartext report directly to the buyer and seller. In our experiments with the current implementation, we also observed additional messages with associated privacy risks depending on the server and browser implementation. For example, when we set a "No-Store" cache policy for the buyer server, the browser sends a request to retrieve the buyer's bidding code via the url specified in \keyword{biddingLogicURL} between Step 4 and Step 5 (shown as the dashed line in \autoref{fig:design}). This is a privacy concern as the buyer server can log information of this cross-site request, such as timestamp, IP address, and user-agent headers. 

Although ensuring the deployed protocol matches some specified protocol that has been analyzed to satisfy desired properties is essential, \fledge is not yet at the stage where the actual protocol is clearly specified. Since our goal is to understand fundamental issues with the intended \fledge protocol, we conduct our analysis based on our best interpretation of the plans for \fledge, assuming that proposed privacy mechanisms (like the $k$-anonymity server) will be implemented and that this and other implementation issues will be fixed in the eventual \fledge\ implementation. Our focus is on identifying issues in the intended \fledge\ protocol as designed instead of the current implementation in the \FOT.

\subsection{Attaching an Interest Group} \label{ig}

To attach an interest group to the user's browser (\autoref{fig:design}, Step 2), the buyer's script in the browser calls the \keyword{joinAdInterestGroup($\cdot$)} API with two parameters: \keyword{interestGroup} and \keyword{time}. 
The \keyword{time} parameter to the \keyword{joinAdInterestGroup($\cdot$)} API is a number specifying the duration of the membership in seconds. The maximum value \fledge allows for \keyword{time} is 30 days (2,592,000 seconds). 
%After calling this API, the interest group object will be stored in the browser for specified time. 
%
The \keyword{interestGroup} is a JSON object that includes: (1) the interest name, an arbitrary string selected by the buyer site; (2) the owner of the interest group which is typically the buyer site's origin; (3) a bidding URL; (4) an update URL, which the browser requests once daily to replaces fields (except the interest group name and owner) in the original interest group object; and (5) an \keyword{ad} field that can contain arbitrary ad metadata. The bidding URL provides the buyer's code will be run in a browser worklet to generate bids and report auction results. To avoid side channel leaks, this code should be requested periodically and cached in the browser (the current \FOT implementation does not always do this, as discussed in \autoref{sec:protocol-overview}). 

\subsection{Initiating an Auction} \label{auction-config}

To start an auction (\autoref{fig:design}, Step 4), the seller's script in the browser calls \keyword{runAdAuction($\cdot$)}, passing in \keyword{auctionConfig}, a JSON object that contains information including the seller domain URL and a decision URL. Similar to the bidding URL, it points to seller's code that is later used in the browser to score different bids and report auction results. The \keyword{auctionConfig} contains other fields such as \keyword{auctionSignals}, which will be passed to all participating buyers during the auction.
%, and \keyword{perBuyerSignals}, which enables the seller to send distinct messages to individual buyers. 
%After calling this API, the \keyword{auctionConfig} object will be stored in the browser.  

\subsection{Running the Auction in the Browser}\label{sec:runningauction}

After the seller's script initiates the auction by calling \keyword{run\-Ad\-Auc\-tion($\cdot$)} (\autoref{fig:design}, Step 4), the browser creates worklets running code provided by the buyers and sellers. Functions running inside the worklets provide controlled communication channels through their inputs and outputs, and the browser connects those inputs and outputs across the worklets. To prepare the execution environment of the auction, the browser first iterates through all the interest groups associated with the user and creates one bidding worklet for each interest group the seller allows to participate in the auction (specified in the \keyword{auctionConfig} object). It also creates an auction worklet for the seller. Then, the browser obtains the code for each buyer involved in the auction (which should already be cached in the browser) from the \keyword{biddingLogicURL} field in each interest group and gets the seller's code from the \keyword{decisionLogicURL} field in seller's \keyword{auctionConfig} object. The code for each buyer will run in a corresponding bidding worklet to provide bids for each interest group and the seller's code will run in the auction worklet to score these bids. The ad with the highest score is selected as the auction winner to be displayed in the browser.

\subsection{Aggregate Reporting Service} 
\label{aggregate-service}

In the \FOT, event-level reporting happens right after an auction ends, as the winning buyer and the seller can send arbitrary information as arguments to \keyword{reportWin($\cdot$)} and \keyword{reportResult($\cdot$)} in the form of a URL parameter.
This poses obvious and severe privacy risks. For example, it allows both parties to determine when a user visited the seller's site. The buyer could also include a tracking ID in the \keyword{interestGroup} object and send it to the seller via the ad metadata in the \keyword{generateBid($\cdot$)} function. Similarly, the seller could include a tracking ID in the \keyword{auctionConfig} object when initiating the auction and transmit it to the buyer through the \keyword{sellerSignals} in the \keyword{reportResult($\cdot$)} function. Through this method, the buyer and seller can track a user by exchanging tracking IDs through the reporting URL.

The proposed \fledge\ design avoids the most obvious privacy violations by encrypting event-level reports and waiting a random time delay before they are sent. The Trusted Aggregation Service aims to allow participants of \fledge auctions to create summary reports that can be used to understand ad placements without compromising individual users (Step 7 in \autoref{fig:design}). To maintain privacy, the aggregation service code must be audited and approved before deployment and operate in a TEE in a public cloud platform. 

As depicted in \autoref{fig:design}, only the winning bidder learns about the auction. Buyers can also implement a \keyword{reportLoss($\cdot$)} function that provides information to the buyer when they participate but lose an auction. The details of what information is passed to this have not yet been determined.  The current plan is to support unencrypted event-level reporting in the \keyword{reportWin($\cdot$)} function until at least 2026, while using the the \keyword{privateAggregation($\cdot$)} API for losing buyers, but the details of this plan have not yet been disclosed~\cite{extendedreporting}. For our analysis, we assume the proposed design (post-2026) in which only encrypted reports are sent back to the buyer or seller and all semantic information about auction results has to be obtained through the aggregation server.

After an auction ends, the buyer's code can generate event-level reports in the form of key-value pairs $(k, v)$ by calling the \keyword{privateAggregation($\cdot$)} API within the worklet. The bucket key $k$ is a \keyword{BigInt} with maximum length of 128 bit and the value key $v$ is a 32-bit integer. The browser then generates an encrypted event-level report by encrypting the $(k, v)$ pair. Finally, the browser sends this encrypted event-level report to the buyer and seller's server after some random delay time up to an hour. 

In order to prevent adversaries from using the reports to leak too much information, \fledge implements restrictions on the API calls in the reporting function. When the buyer or seller calls the \keyword{privateAggregation($\cdot$)} API in the worklet function, the browser limits the total value that can be contributed to bucket keys. The sum of all contributions across all buckets, i.e., the total value of all the contributions, may not exceed $2^{16}$. If this limit is exceeded, future contributions will be dropped without notification, meaning no further encrypted event-level reports can be sent to the API caller's server once the budget is depleted. The contribution limit will be reset over a rolling 10-minute window for each buyer or seller site with a daily cap of $2^{20}$.

Later, when buyers and sellers query the aggregation servers, they send a collection of encrypted reports along with a list of bucket keys to be included in the aggregated report. The aggregation service sums the value for each bucket across all provided encrypted reports, add noise to provide a differential privacy bound, and returns an aggregated report as a histogram with a noised sum for each bucket. All bucket keys in the query will appear in the summary report; if a key does not appear in any of the encrypted reports, it will still be included as zero plus noise value in the aggregated report.

%To prevent adversaries from de-noising the contribution value, such as by repeatedly querying the same encrypted report in multiple batches, 
The coordinator in Trusted Aggregation Service limits the number of queries each report, and \fledge plans to enforce a non-duplicate rule, only allowing one query per event-level report. This means each encrypted report can only appear once within a batch and cannot contribute to more than one aggregated report.

\subsection{Checking for \emph{k}-anonymity} \label{ssec:k-anon}

The \fledge design includes $k$-anonymity checks with the aim of preventing buyers from using interest groups to track users~\cite{kanonplan}. The $k$-anonymity \cite{samarati1998protecting} privacy notion aims to safeguard the identities of individuals within a dataset by grouping them into clusters of at least $k$ individuals with similar attributes. 
%
%The \FOT\ does not implement any $k$-anonymity checks, but 
\fledge plans to enforce $k$-anonymity with $k=50$ over a 30-day period with hourly updates. 

%There are two places where $k$-anonymity checks are done~: (1) to determine whether an ad can participate in an auction, and (2) to limit whether the name of an interest group appears in a report. 
The $k$-anonymity tracking is done using the $k$-anonymity server, with \keyword{Join($\cdot$)} API calls used to record group membership and \keyword{Query($\cdot$)} calls used to check if there are at least $k$ members in the group associated with an object. These calls are shown in \autoref{fig:design}, but the actual calls are made outside of the critical path of an ad auction. The browser periodically reports new objects, which are created when a buyer calls the \keyword{joinAdInterestGroup($\cdot$)}, via \keyword{Join($\cdot$)} calls. Similarly, the browser periodically sends \keyword{Query($\cdot$)} requests to the $k$-anonymity server for relevant objects, caching the results for use during in-browser auctions. 

The \keyword{Join($\cdot$)} call has three parameters: a browser identifier \keyword{b}, a type \keyword{t}, and an object represented by a hash \keyword{s}. The browser identifier \keyword{b} is a $j$-bit identifier, where $8 \leq j \leq 16$. The space of browser identifiers is large enough to provide a sufficiently tight lower bound on the number of different browsers in a group, but these identifiers are not unique. Many browsers will share the same identifier, which mitigates the privacy risks associated with the identifier. Since the $k$-anonymity check is performed on both interest group name and ad creative URLs, \keyword{t} specifies the type of the object. % and \keyword{s} is the hash of each checking object. 
At the $k$-anonymity server receiving the \keyword{Join($\cdot$)}, the browser's identifier \keyword{b} is inserted into the set of browser identifiers associated with the object \keyword{s}. The size of this set is what determines if a $k$-anonymity check is passed. 

The \keyword{Query($\cdot$)} call has two parameters: a type \keyword{t} and an object represented by a hash \keyword{s}. When the browser makes a \keyword{Query($\cdot$)} call, it checks the number of browser identifiers associated with the object \keyword{s} and responds with a Boolean value indicating whether the $k$-anonymity check has passed (that is, the group associated with \keyword{s} has at least $k$ members). 

The $k$-anonymity check is applied in two places: 1) during the auction phase---where an ad must pass the $k$-anonymity check to participate in the auction; and 2) at the reporting time, to determine if the interest group name is included during report generation. The difference between these two checks is whether or not the interest group name is included in the checking object. For the first check, the interest group name is not included, and the object is a tuple consisting of the interest group owner's URL, the bidding script's URL, the creative's URL, and the ad's size. The reason the interest group name is not included in the first object to check is to allow advertisers implement various bidding strategies (e.g. multiple small-size interest groups can share some ads). For the second check, the interest group name is also included, meaning that there are interest groups that pass the first check to be eligible for the auction, but not the second one. This is intended to prevent advertisers from using information in the interest group name to generate report if the interest group has less than $k$ users.

Although the $k$-anonymity check is designed to prevent tracking individuals, in \autoref{sec:circumvent-k-anon} we show how the $k$-anonymity checks are insufficient for preventing high confidence large-scale surveillance.

\section{Attack Overview} \label{sec:attackoverview}

In the current online advertising ecosystem, third-party cookies allow websites to track users across sites and build user profiles. This information is then used by buyers and sellers to tailor personalized ads to users. The goal of \fledge is to support targeted advertising without third-party cookies or a centralized data collector. Therefore, it is worth investigating how well \fledge achieves this goal of preventing user tracking.

To limit our scope, we only consider the goal of protecting user privacy from buyers and sellers operating in the ad ecosystem and focus on the ability of an adversary to link two requests. There are numerous other security and privacy aspects of \fledge, including preventing buyers and sellers from manipulating the ad auctions to their benefit, protecting the confidentiality of buyers from other buyers and sellers, and other ways information about user interests can be leaked to buyers and sellers. These are important issues that merit consideration, but are outside the scope of this analysis which only considers user tracking through linking requests.

\subsection{Threat Model}\label{sec:threat}

To assess the effectiveness of \fledge in preventing tracking, we consider a threat model where adversaries who can control buyer domains and seller web servers attempt to gather information about users. The adversary's goal is to link requests made to two different sites to the same user by using information provided through \fledge reports. 

We assume that other currently available ways to link user requests will not be available, so the adversary cannot identify a user through a stable IP address or digital fingerprinting. Without masking the IP address or protections against fingerprinting, most users today can be uniquely identified by their visit directly \cite{karpukhin2022anonymization, mishra2020don}, so there would be no additional linking risk from \fledge\ in situations where the two requests can already be linked. In a future where fingerprinting protections are deployed and third-party cookies are eliminated, it becomes difficult to track users and link requests directly. Hence, the focus of our analysis is to understand how much these risks increase over when \fledge\ is deployed over the (hopeful future) baseline where user tracking would otherwise be infeasible. 
%With \fledge enabled, websites can potentially acquire user information across different sessions through the \fledge\ protocol messages and the information released through aggregated reports.

%, even if their digital fingerprint changes, by syncing their tracking IDs after an auction concludes. It is worth noting that \fledge is only responsible for generating summary reports for auction outcomes, leaving the actual billing process to buyers and sellers. Thus, buyers and sellers may use \fledge as a protocol for linking cross-site requests instead of for remarketing purposes with limited cost. 

We focus our study on the design of \fledge\ and the information channels provided through auction reports, so further assume the browser can be trusted and that everything works as intended. This requires that: 
\begin{itemize}
  \item Dedicated worklets created by the browser are isolated without access to outside network or storage, and there is no way for code running in other worklets to exploit side-channel attacks to infer information from these communication channels. 
  \item The TEEs used to execute the K/V servers and Aggregation servers (\autoref{aggregate-service}) operate securely as trusted execution environments without leaking and information, and the attestation and management of the attestation keys is all correct and invulnerable.
  \item The process used to audit the code for the K/V and Aggregation servers is sound and ensures that any code that passes this process cannot violate the required properties. 
\end{itemize}

All of these are strong assumptions and difficult to achieve in practice. 
%, but for our analysis we assume that they can be achieved. 
Even in cooperative settings, implementing differential privacy mechanisms correctly (as is required for the aggregate reports) is challenging~\cite{ding2018detecting, kifer2020guidelines, TTSS+22, jin2022we} and analyzing software for security vulnerabilities is challenging and can rarely be done completely~\cite{goseva2015capability, buildbreak2020,votipka2020understanding}; in adversarial settings where the code author may be deliberately making their code hard to analyze and to hide a prohibited behavior in it, relying on perfect audits requires a leap of faith~\cite{bannet2004hack}. There are many subtle ways a program can be designed to intentionally leak data~\cite{reardon201950}, and no known method, even with source code available, to ensure all leaks are detected.

\subsection{Attack Steps}\label{sub:attacksteps}
To execute the attack, the adversary needs to control at least one primary site where visitors reveal their identities, and a sensitive secondary site where users expect to be anonymous. The adversary's goal is to link requests to the sensitive site to requests to the primary site, and thereby learn the identity of an anonymous user on the sensitive site. In addition, the adversary may control some other domains, which are enrolled as $n$ buyers. 

The general attack process unfolds over the following four steps:
\begin{enumerate}[leftmargin=4em,itemsep=1.5ex,labelwidth=1cm,labelindent=0.5cm,label=\bfseries Step \arabic*.]%,align=left]
\setcounter{enumi}{0}
\item Users visit the primary site where the adversary associates the user’s browser with $n$ interest groups, each representing a different buyer controlled by the adversary.
\item Before the interest groups expire, a user navigates to the secondary site, which is a sensitive site where the user expected to be anonymous. The adversary controls an ad auction on the secondary site, and lets each of the $n$ colluding buyers win one of $n$ separate auctions. 
\item After winning the auction, each colluding buyer generates encrypted reports using identical bucket-value pairs, which are then sent to the adversary.
\item The adversary submits all the encrypted reports with selected bucket keys to the Aggregation Service and received the aggregated output. By analyzing the aggregated output, the adversary links user visits across the primary and sensitive sites to identify users on the sensitive site.
\end{enumerate}

\autoref{sec:analysis} explains how an adversary can execute the attack in three different scenarios: linking a single targeted user, linking one out of many users, and conducting large-scale surveillance.

%Further details on the variations of this attack process, including how it scales with different numbers of users and the necessary size of $n$, are in .

\subsection{Attack Feasibility}
As outlined in the attack steps, the attack which enables linking is only possible in settings where a user makes requests to both a primary site and a sensitive secondary site, both of which are controlled by the adversary who wants to link the requests. In \fledge, the duration that an interest group can remain active in a user's browser is capped at 30 days. Consequently, the feasibility of a linking attack is contingent upon the user visiting both sites within this 30-day window.

Effectiveness of the attack also depends on an adversary's ability to control multiple buyer domains, and Google controls access to the APIs needed to participate in ad auctions. 
%\shortsectionnp{{\em How hard is it to control many buyer sites?}}
As various APIs from Google's Privacy Sandbox begin to reach general availability, Google has outlined plans to implement a verification process for entities accessing these APIs. Google initiated the enforcement of enrollment starting in mid-October 2023, with the release of Chrome 118 Stable. To ensure auditable transparency, the enrollment information related to each company will be publicly accessible. This verification procedure is designed to mitigate API misuse such as preventing one developer from impersonating another and restricting their access to the APIs. However, the underlying business model of selling ads to many buyers around the world is incompatible with a restrictive or costly verification procedure. 
Enrolling as a buyer or seller to access these APIs only requires basic business contact information, a D-U-N-S number for the organization (which can be obtained for free by providing some information in a web form)~\cite{duns}, and the necessary input for API or server configurations~\cite{enrollment}. Consequently, it does not seem unreasonable for a motivated adversary to be able to enroll many buyers. Our analysis in  \autoref{ss:large-pool-target} shows that 200 buyers is sufficient for large-scale surveillance with high confidence based on just a single visit to the secondary site. 
\section{Data Leakage Analysis} \label{sec:analysis}

A primary motivation of the push to eliminate third-party cookies is to make linking user behaviors across websites infeasible, or at least expensive enough for an adversary in order to disrupt the most extensive user tracking. 
Hence, we focus our analysis on how well the \fledge\ design maintains the unlinkability goals that underlie the push to eliminate third-party cookies. We consider three different scenarios based on the number of candidate individuals to link: \autoref{ss:single-targeted-user} is the simplest case where the adversary has a single target individual in mind who visits a first server and wants to determine if they are linked with a request to a second server; \autoref{ss:multiple-targeted-user} considers an adversary who wants to link any one of many candidates across requests; and \autoref{ss:large-pool-target} analyzes an adversary who want to perform mass surveillance by linking requests from a large pool of candidates across two sites.

The aggregate reporting service (\autoref{aggregate-service}) and the $k$-anonymity check (\autoref{ssec:k-anon}) are two main features in the \fledge\ design intended to prevent request linking. For clarity of presentation, we do not include the $k$-anonymity check in this section, but show in \autoref{sec:circumvent-k-anon} how the $k$-anonymity checks can be circumvented.

\subsection{Scenario 1: Linking a Single Targeted User} \label{ss:single-targeted-user}
Consider a simple scenario where an adversary controls two sites---one buyer site and one seller site. We assume that the targeted user visits a primary site, and some time later, visits the secondary site. The adversary's goal is to link these two requests. This models the scenario where, for example, the primary site is a non-sensitive site such as a shopping or news site that the user visits without concealing their identity and the secondary site is a politically sensitive site which a user visits expecting anonymity. In the \fledge\ protocol, the primary site acts as an ad buyer, and the secondary site as an ad seller.

When the targeted user visits the primary site, the site can associate a user-specific interest group with the user.\footnote{The $k$-anonymity checks should prevent this interest group from participating in the auction, but as we discuss in \autoref{sec:circumvent-k-anon}, they can be circumvented.}
Later, when the user visits the secondary site, that site sets up an auction that will be won by the primary site as the buyer. This causes the buyer's worklet running in the browser to generate an encrypted report in the \keyword{reportWin($\cdot$)} function (Step 7, \autoref{fig:auction-process}), which is sent to the buyer's  site. Therefore, once the primary site receives this encrypted report, without needing to decrypt it, the adversary can already determine (with certainty) that the targeted user visited the secondary site. The protocol imposes a random delay of up to an hour before the report is transmitted, so the buyer will not learn the exact time of the visit, but will know that the specific targeted user visited the secondary site sometime within a hour of the time when they receive the report.

This attack illustrates the danger of covert channels in a setting where adversaries have the ability to run their own code in an environment with access to sensitive information, and have a communication channel back to receive results from this code. It violates the unlinkability goals of \fledge, but only in a very limited way. Nevertheless, such a simple attack may already be a serious privacy risk in some scenarios such as when an oppressive government has a desire to gather evidence against a suspected dissident, but its scale is limited in that within a given time period only a single, predetermined victim identity can be linked. In the next scenario, we consider a more scalable linking attack.

\subsection{Scenario 2: Linking One of Many Users} \label{ss:multiple-targeted-user}

We consider a simple but realistic scenario where in addition to controlling the primary and secondary websites, the adversary also controls some additional ad buyer sites. Instead of just linking a single known user as in the previous scenario, now the adversary has  a list of $k$ candidate users (which could be everyone who visits the primary site). We assume that some of the candidate users visit the primary site and subsequently visit the secondary site. The adversary's goal is to link these two requests to identify with confidence which of the candidate users have visited the secondary (sensitive) site.

When a user visits the primary site, it can call the \keyword{joinAd\-Interest\-Group($\cdot$)} API on behalf of all $n$ colluding ad buyer sites (one of which is the primary site), assigning a user identifier (UID) included in the interest group name. This allows each buyer site to associate a user-specific interest group with the user. Later, when the user visits the secondary site, it offers $n$ ad spaces, each with an associated auction that is designed to be won by a different one of the $n$ buyers.\footnote{There is no apparent limit to the number of ad spaces that can be sold on a single webpage visit. We have tested the\fledge protocol with up to 200 buyers.} 

As outlined in \autoref{aggregate-service}, upon winning an auction, buyer worklets running in the browser can generate encrypted reports in the form of key--value pairs that are sent to the buyer's server. In this case, in the \keyword{reportWin($\cdot$)} function, all buyers use the same user identifier (UID) recorded in the interest group name as the key and the full buyer-sensitivity budget as the reporting value.
Consequently, within an hour of the auction concluding, each of the $n$ winning buyers receives an encrypted report with value $\ell_1$ recorded in the same UID bucket. 

Since these encrypted reports will be sent out of order within one hour after the auction ends, when any one of the colluding buyer sites receives the first encrypted report it can notify the secondary site which can update the \keyword{auctionConfig} to prevent this set of colluding buyers from participating in any further auctions until all $n$ reports have been received. In this way, when the adversary collects the batch of encrypted reports from the $n$ buyers within an hour, it is certain that only one user out of $k$ target users visited the tracked site. This simplification makes the analysis easier and enables high confidence for in the one target user identified. In \autoref{ss:large-pool-target} we describe a more efficient scheme for tracking large numbers of users.

Once all $n$ reports have been received, the adversary queries the aggregation service using the set of UIDs corresponding to the list of target users. 

\shortsection{Query Semantics}
Let's denote $Q$ as a query function executed by the buyer to the aggregation service, $R$ as a set of encrypted reports chosen for a specific query, and $B$ as a list of bucket keys chosen for a specific query. Then we can represent a query and its response as $S \leftarrow Q(R, B)$, where $S_i = \Sigma_{r \in R} B^r_i + \mathit{Laplace}( 0, \frac{l_1}{\epsilon})$, where $\epsilon$ is the privacy loss budget. When the browser enforces a maximum of $l_1$ impact for each user visit on any site over 10 minutes, the sensitivity of a single report is $l_1$, so this provides $\epsilon$-Differential Privacy according to the Laplace Mechanism ~\cite{dwork2014algorithmic}.

The aggregation service outputs a vector of length $|B|$, where each value is a noised value. For the user who visited the tracked site, the aggregated output value of that UID will be $n \cdot \ell_1+\mathit{noise}$; for all of the other UIDs the value will be $0+\mathit{noise}$, where each noise value is sampled independently from the Laplace distribution to satisfy the $\epsilon$-DP guarantee (assuming sensitivity $\ell_1$, which is now effectively violated by the collusion). A simple attack just accuses the user with the UID associated with the highest value in the aggregate histogram.   

\begin{figure}[tb]
    \centering
    \includegraphics[width=0.78\columnwidth]
    {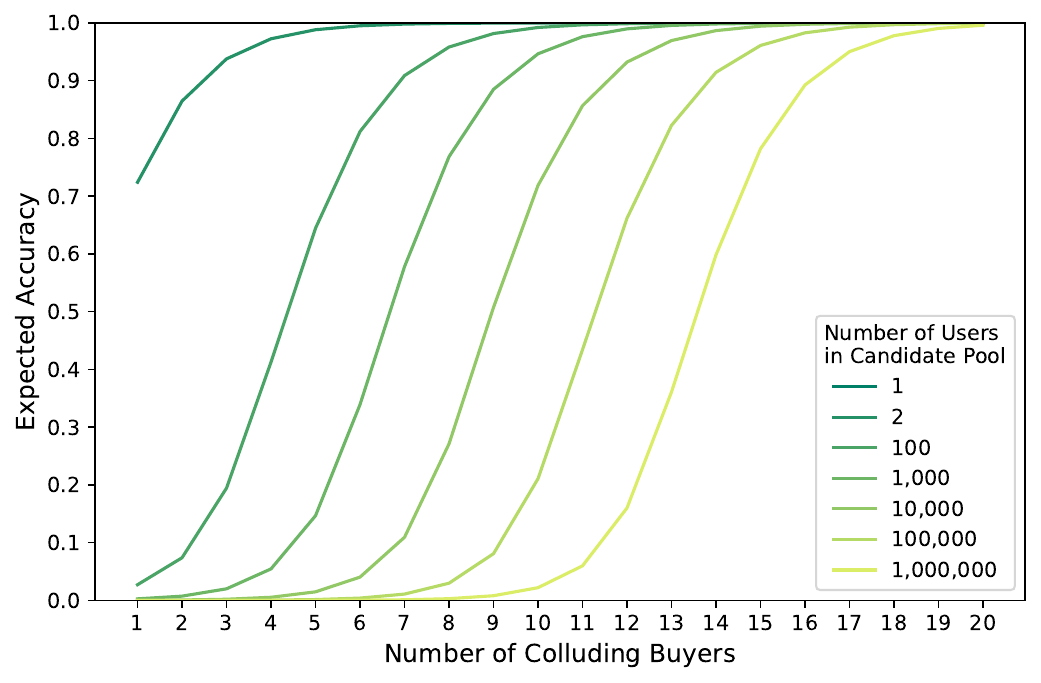}
    \caption{Expected accuracy for predicting the targeted user's presence on the secondary site ($\epsilon = 1$).}
    \label{fig:accuracy}
\end{figure} 

\shortsection{Effectiveness} 
To evaluate the effectiveness of this attack, we present a theorem to quantify the expected accuracy of adversary's prediction algorithm with respect to the number of target users, the number of colluding buyers, and the privacy loss parameter. 

\begin{theorem} \label{theorem-accuracy} \rm
The number of target users is $u$, $n$ is the number of colluding buyers controlled by the secondary site, and $\epsilon$ is the privacy loss parameter. Each $Y_z$ is independently sampled from a Laplace distribution, $\mathit{Laplace}(0, \frac{1}{\epsilon})$. Given an aggregated report consisting of noisy outputs $\vec{x} = [x_1, x_2, \ldots, x_u]$, and where for one specific $j \in u$, $x_j = n + Y_j$ and for all other $i \in u$, $i \neq j$, $x_i = Y_i$, there exists an algorithm that given $\vec{x}$ can predict $j$ with expected accuracy:
\[
    \mathit{Accuracy} = \int_{-\infty}^{\infty}f_{\mathit{Laplace}}(Y_j) \cdot F_{\mathit{Laplace}}(n+Y_j)^{u-1}dY_j
\]

\end{theorem}

Here, $f_{\mathit{Laplace}}(x)$ and $F_{\mathit{Laplace}}(x)$represent the probability density function and the cumulative distribution function of $\mathit{Laplace}(0, \frac{1}{\epsilon})$. See \autoref{theorem-explanation} for a proof of the theorem and an explanation of numerical method we use to approximate the value of the integral.

\autoref{fig:accuracy} shows the relationship between the number of colluding buyers and the accuracy of the adversary's prediction algorithm. We show the results for when the privacy loss parameter $\epsilon$ is set to 1. With the \fledge default setting of $\epsilon = 10$, the expected accuracy exceeds 99\% with only two colluding buyers with 1 million targeted users, so we analyze with a lower privacy loss parameter of $\epsilon = 1$. As the aggregation service does not pose a limit on the length of bucket list in the query and the bucket key has a maximum of 128 bits, the number of candidate users $u$ can be large. For reference, it takes around 35 seconds for the Aggregation Service in local set up to process a bucket key list with length of 1 million entries \cite{localtesting}. Thus, identifying one user out of millions of potential candidates appears to be realistic.

As the number of reports (which is the number of colluding buyer sites controlled by the adversary) aggregated increases, the adversary quickly reaches a high expected accuracy on identifying which user on the target list visited the tracked site. With 13 colluders (the primary site and 12 additional buyers), the expected accuracy in prediction of this simple attack exceeds 99 percent when the target user list has 1,000 users. Even when there are one million target users, an adversary with at least 15 colluding buyer sites can identify a user who visits the secondary site with high confidence. This is alarming---obtaining buyer identities is unlikely to be difficult and the economics of the system require that it be relatively easy to set up new entities as ad buyers.

\shortsection{Enhancements}
The contribution limit for a single report is set at $2^{16}$ which is used as the sensitivity in the differential privacy mechanism. However, this limit is reset within a rolling 10-minute window for each buyer or tracked site, with a daily cap of $2^{20}$. This means that if the target user visits the secondary site multiple times throughout the day, the adversary could potentially generate up to $16 \cdot n$ reports per day for that single target user. This increased volume of reports can significantly improve the expected accuracy in prediction in settings where only a few colluding buyers are available. Furthermore, instead of using full contribution budget when a target user visits the tracked site, the adversary can allocate the budget in other ways to track more than one user at a time, which we discuss next.

\subsection{Scenario 3: Mass Surveillance} \label{ss:large-pool-target}

The attack in the previous scenario can identify a single user from a large candidate pool, but does not demonstrate the risks of mass surveillance that are the primary motivation for eliminating third-party cookies. In this scenario, we analyze the potential for mass surveillance after third-party cookies are eliminated and \fledge\ is deployed. Here, we consider the same scenario as in \autoref{ss:multiple-targeted-user}, but instead of linking one user with $n$ reports per query to the Aggregation Service, the adversary collects $n \cdot u$ reports from $n$ colluding buyers corresponding to $u$ candidate user visits, queries the aggregation service with a fixed set of bucket keys, and infers the presence of many users (approaching the number of visitors, $u$) out of the large candidate pool based on the output. 

The main idea behind the attack is instead of using each bucket to represent one user, we use the buckets as a Bloom filter to enhance tracking capabilities. The process of reporting and querying is a straightforward application of the Bloom filter \cite{bloom1970space}, modified to account for noise in the aggregate reports. In a standard Bloom filter, set membership is recorded using a vector $B$ of size $m$. There are $a$ independent hash functions, $h_1, \ldots, h_a$, and when an element $z$ is added to the set the values of each corresponding bit $h_i(z) \mathbin{\%} m$ is set to 1. An element $z$ is predicted to be in the set if $\Sigma_{i\in{1 \ldots a}} B[h_i(z) \mathbin{\%} m] = i$. This provides a guarantee of no false negatives, and with appropriate parameters can achieve a very low false positive rate. 

To adapt this approach to the \fledge\ setting the we need to account for the noise added to the aggregate reports. The protocol design is similar to the one from Scenario~2, except now each colluding buyer computes the hash the UID with $a$ distinct hash functions $h_1 \ldots h_a$, and increments each $B[h_i(z) \mathbin{\%} m]$ by  $\ell_1 / a$, dividing the sensitivity budget evenly across the buckets. Thus, each buyer will call the \keyword{privateAggregation($\cdot$)} API $a$ times to send key--value pairs as ($B[h_i(z) \mathbin{\%} m]$, $\frac{\ell_1}{a}$) where $i \in a$.  When there are collisions, the same bucket value is incremented multiple times. 

To infer around $u$ visitors, when the adversary has received $n \cdot u$  reports, they notify the secondary site to suspend letting colluding buyers win any further auctions for the next hour, and then wait for up to an hour to receive all reports to account for the random delays in sending reports. Modulo race conditions, this minimizes the risk that the collected reports will include mismatches (different visits reported across the $n$ colluding buyers). In practice, mismatched reports would just mean that for some visits fewer than $n$ colluding buyer reports have been received when the aggregation is done, so an adversary may prefer to just collect reports continuously and perform aggregation periodically instead of suspending collection for an hour between every aggregation period. For our analysis, we keep things simple and assume a complete set of reports.

The adversary sends all the reports received to the Aggregation Service with a query for bucket keys $0, \ldots, m-1$. 
This results in an $m$-bit array, $B_{\mathrm{report}}$, with the noisy histogram output. The adversary's goal is to determine which of the candidate users visited the secondary site during the tracking period. For each target user in the candidate pool, the adversary performs the same hashing process on each $UID_{\mathrm{target}}$, which results in another $m$-bit array $B_{\mathrm{target}}$. For each index with a non-zero value in $B_{\mathrm{target}}$, the adversary estimates the probability that the noisy value in $B_{\mathrm{report}}$ at that index corresponds to a non-zero true value. Then, they take the product of all these probabilities to calculate the likelihood that all of these bit indices in $B_{\mathrm{target}}$ are non-zero, as would be the case if $UID_{\mathrm{target}}$ visited the tracked site. Lastly, the adversary sorts these likelihood scores and accuses some users with highest scores.

\shortsection{Analysis}
To evaluate the effectiveness of this attack, we measure the number of visitors linked and the positive predictive value (PPV) through simulations and explain how we select parameters, such as the fixed domain of bucket keys $m$ to maximize detection with high PPV.
The false positive rate for a standard Bloom filter with large $m$ and small $a$ is extremely close to the theoretical bound $(1-(1-\frac{1}{m})^{au})^a$ \cite{bose2008false}, where $m$ is the length of Bloom filter, $a$ is the number of hash functions, and $u$ is the number of elements stored in the filter. \fledge limits the number of contributions in a single report in the \keyword{reportWin($\cdot$)} function to 20 \cite{constraintonk}, so the number of hash functions can be up to $a=20$ which is the value we use. We analyze the case where $u = 10,000$, which means the adversary aims to track 10,000 users who visit the tracked site at least once during each surveillance period. To minimize the (non-noised) false positive rate to $0.01\%$ with $a=20$ and $u=10,000$, the adversary uses a Bloom filter with $m = 201,000$ bits.

%As bucket keys can be an arbitrary 128-bit BigInt, the length of the Bloom filter can be up to $m = 2^{128}$ bits (since we need to query all buckets in the aggregation protocol, though, . 

After receiving the noisy outputs from the Aggregation Service, the adversary's task is to predict which users visited the tracked secondary website. The adversary does this by examining the returned vector of $m$ noisy values and checks the $k$ indices corresponding to each user. For each user, these $k$ noisy values in $B_{\mathrm{report}}$ are at the non-zero positions of $B_{\mathrm{target}}$. We denote this as $\vec{x} = [x_1, \cdots, x_a]$, where each $x_i$ is the output after adding noise at that position, $x_i = Y_i + c_i$. Here, $Y_i$ represents a random variable drawn from a Laplace distribution with parameters $(0, \frac{\ell_1}{\epsilon})$, and $c_i$ represents the original value contained in the encrypted report. 
%(which is either 0 or a low multiple of $n \cdot \ell_1 / k$).

In cases where there are $n$ colluding buyers and the user has indeed visited the secondary site, it holds that $c \geq n \cdot \ell_1 / a$. In the case where this user visited once and there are no hash collisions, it would be exactly the minimum value $c = n \cdot \ell_1/a$. Collisions only increase the probability of detection, so we ignore than in the rest of this analysis.

In this binary context, the adversary wants to compute the probability that $c_i$ is non-zero ($c_i = n \cdot \ell_1 / a$) for each observed $x_i$: 
\begin{equation}
P(c_i = n \cdot \frac{\ell_1}{a} \; | \; x_i) = \frac{f_{\mathit{Laplace}}(x_i-c_i)}{f_{\mathit{Laplace}}(x_i) + f_{\mathit{Laplace}}(x_i-c_i)}
\end{equation}
Here, $f_{\mathit{Laplace}}(x)$ represents the probability density function of the Laplace distribution $(0, \ell_1 / \epsilon)$, reflecting the probability that the bit at index $x_i$ has been set in the Bloom filter (so the denominator considers the only two cases in this binary setting---either $c_i = 0$ or $c_i = n \cdot \ell_1 / a$.

The adversary's goal is to distinguish between two hypotheses: the null hypothesis, $H_0$, is that the user did not visit the tracked site; the alternative hypothesis $H_1$ posits that the user visited the tracked site at least once. Given the vector $\vec{x}$, the adversary can calculate the likelihood that the original values at all these indices are $c=n \cdot \ell_1 / k$, denoted as $L(H_1 \; | \; \vec{x})$:
%and compute the conjunctive probability as:
\begin{equation}
L(H_1 \; | \; \vec{x}) = \prod_{i=1}^{i=a}{P(c = n \cdot \frac{\ell_1}{a} \; | \; x_i)}
\end{equation}
Finally, the adversary sorts all the users by their likelihood score, and accuses the users with highest scores. The adversary has a good estimate of the actual number of visitors to the secondary site, $u$, based on the number of reports received, which is equal to $u \cdot n$. The number of accusations can be varied to trade-off between false positives (invalid accusations) and false negatives (not accusing a user who did visit the secondary site).

\shortsection{Results}
\autoref{tab:epsilon-top-k} shows the smallest number of colluding buyers the adversary needs to control to achieve a PPV over 0.99 with respect to different privacy loss budget $\epsilon$ and the number of accusations.  In the simulations, we randomly select 10,000 users to visit the secondary website from a candidate pool of one million users who visited the primary site. We compute the average PPV of the attack across five simulations with respect to different numbers of accusations with varying numbers of colluding buyers. 

While the value of $\epsilon$ has not yet been determined in \fledge, the currently available Aggregation Service for the Private Aggregation API lets developers select an $\epsilon$ up to 64, with a default value $\epsilon = 10$ \cite{localtesting}. The number of colluding buyers needed for high confidence accusations scales up as the value of $\epsilon$ decreases or the number of accusation increases. 

With the default privacy budget of $\epsilon=10$, privacy quickly erodes. With a candidate pool of 1 million users, the adversary only needs $5 \pm 0.4$ colluding buyers to confidently accuse 1,000 users out of 10,000 users who visited the tracked site (all results are averages and variances over 5 simulations). If the adversary wants to make 10,000 accusations, with $20.6 \pm 3.0$ colluding buyers, the adversary can on average identify $9913.4 \pm 1.6$ visitors out of 10,000 true visitors (with $86.6 \pm 1.6$ users falsely linked). 
Even with a lower privacy loss budgets of $\epsilon = 1$, the simulated averaged PPV in the prediction of this attack exceeds 99 percent over 10,000 accusations once there are over 200 colluding buyers ($194.6 \pm 4.6$ in the simulations). 
%when 10,000 users out of 1 million users visited the seller's website if the %adversary accuses full 10,000 users with the attack, resulting in $9906.2 \pm 1.9$ true positives and $93.8 \pm 1.9$ false positives across our five simulated experiments. 
This means the adversary can on average identify around 9906 visitors out of 10,000 true visitors with 94 false accusations. 

%The PPV can further increase with more colluding buyers or lower number of accusations. 

\begin{table}[t]
    \centering
    \begin{tabular}{c c c c c c}
        \toprule
        \multicolumn{1}{r}{Number of Accusations:} & 1000 & 5000 & 10,000 \\
        \midrule
        $\epsilon = 1$ & $40.8 \pm 0.2$ & $76.0 \pm 1.6$ & $194.6 \pm 4.6$  \\
        $\epsilon = 3$ & $13.6 \pm 0.6$ & $25.2 \pm 0.9$ & $67.8 \pm 4.6$ \\
        $\epsilon = 5$ & $8.8 \pm 0.2$ & $15.4 \pm 0.6$ & $39.2 \pm 1.4$  \\
        $\epsilon = 7$ & $6.4 \pm 0.2$ & $11.6 \pm 0.6$ & $30.4 \pm 4.2$  \\
        $\epsilon = 10$ & $5.0 \pm 0.4 $ & $8.4 \pm 0.2$ & $20.6 \pm 3.0$  \\
        \bottomrule \\[1ex]
    \end{tabular}
    \caption{Number of colluding buyers needed to achieve above 0.99 PPV.  \rm     For each privacy loss budget $\epsilon$ (up to the default value of $\epsilon = 10$) and the number of accusations, the result in each cell is the number of colluding buyer sites needed to exceed 0.99 PPV, averaged across five simulations. For all cases, there are 10,000 users who visit the secondary site out of 1M candidates.} 
    \label{tab:epsilon-top-k} 
\end{table}

\begin{figure}[tb]
    \centering
   \includegraphics[width=0.85\columnwidth]
    {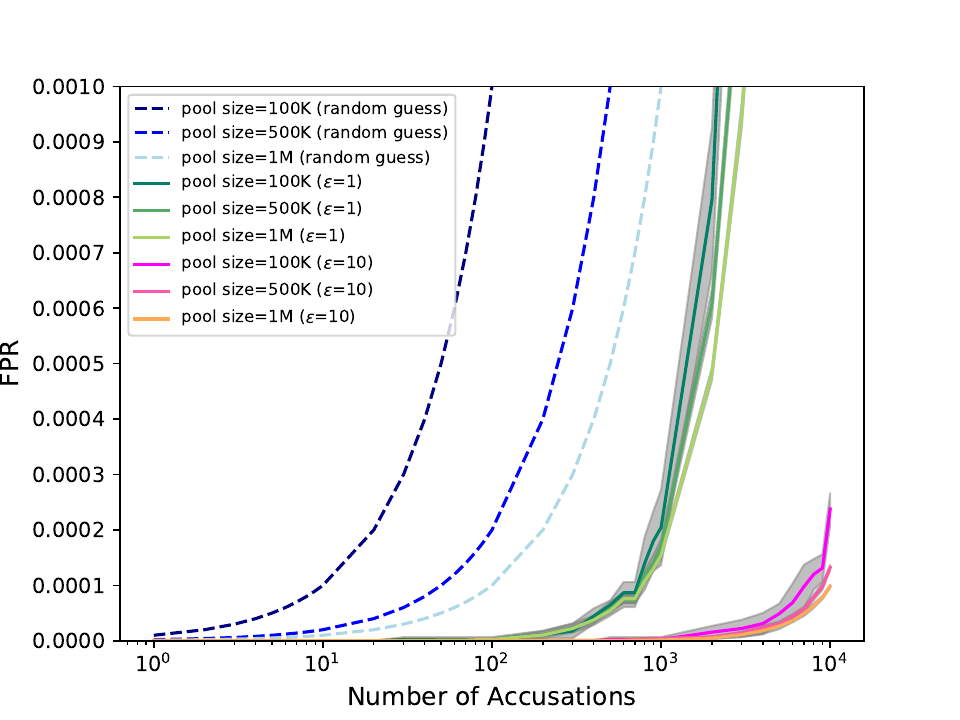}
    \caption{False positive rate as number of accusations varies. \rm Results for setting where adversary controls 20 buyers to predict 10,000 users' presence on the tracked site out of different candidate pools with the default and a reduced privacy loss budget. \rm Results shown are the average over 5 simulation runs. The variance is shown through the shading around the averaged line, where it ranges from 0 to 0.0001.}
    \label{fig:fpr}
\end{figure} 
% \begin{figure*}[h]
%     \centering
%     \begin{subfigure}[b]{0.5\textwidth}
%         \centering
%         \includegraphics[width=0.97\textwidth]
%     {PPV_n5_top_k.pdf}
%         \caption{The number of Colluding Buyers $n = 5$. }\label{fig:simulation-ppv-top-k-n5}
%     \end{subfigure}%
%     ~ 
%     \begin{subfigure}[b]{0.5\textwidth}
%         \centering
%         \includegraphics[width=0.95\textwidth]{PPV_n20_top_k.pdf}
%         \caption{The number of Colluding Buyers $n = 20$.}\label{fig:simulation-ppv-top-k-n20}
%     \end{subfigure}
% \caption{Simulated Positive Predictive Value When Adversary Predicts the 10,000 Users' Presence on the Seller Site ($k$ = 20, $m$ = 201,000) out of Different Candidate Pools with the default and a reduced privacy loss budget. Results shown are the average over 5 simulation runs. The variance is shown through the shading around the averaged line, where it ranges from 0.0009 to 0.4899.}
% \label{fig:simulation-top-k}
% \end{figure*}

\autoref{fig:fpr} shows the relationship between the selection of the number of accusations and the False Positive Rate (FPR) achieved by the adversary's prediction algorithm when the adversary controls 20 buyers for different size candidate pools. 
The FPR of the attack remains very close to 0 (always below 0.00001) when the adversary only makes 100 accusations regardless the candidate pool size up to one million for both $\epsilon = 1$ and $\epsilon = 10$. The FPR becomes unacceptably high for all but the most ruthless accuser at $\epsilon=1$ once the number of accusations exceeds a few hundred, indicating the need for more than 20 colluding buyer sites for a high confidence large-scale surveillance attack.
%closer to random guess if the adversary makes more accusations by performing the attack with only 20 colluding buyers when $\epsilon=1$. In this case, to achieve a low false positive rate, the adversary can either increase the number of colluding buyers or choose to accuse less users. 

%Therefore, if the adversary controls a certain number of product sites and $\epsilon$ is fixed in the final deployment, it is clear that the lesser accusations they make, the more accurate these accusations will be. The adversary can leverage the resources they have (how many product sites they can control) and their tolerance of false positives to find the appropriate parameters of this attack. 

\section{Circumventing \emph{k}-anonymity} \label{sec:circumvent-k-anon}

In the previous section, our attacks in Scenarios 2 and 3 assume the adversary creates a user-specific interest group by including the $UID$ in the interest group name, %while maintaining identical attributes across all other fields within each interest group object, 
and that the seller can extract the $UID$ in the interest group name and convey it to colluding buyers when generating encrypted report. The $k$-anonymity checks in \fledge\ are designed to prevent tracking using user-specific interest groups. 
Recall from \autoref{ssec:k-anon} that there are two types of \emph{k}-anonymity checks. The first check excludes the interest group name when determining ad eligibility for auction wins, thwarting attempts to track users through unique creative URLs. The second check, building upon the first, includes the interest group name in its assessment to decide if the name can be disclosed during report generation, thus preventing re-identification of users via interest group names in auction reports. The attack can easily pass the first $k$-anonymity check by having $k$ users visit the primary site. However, since the interest group name is specific to each user, the second $k$-anonymity check will fail, blocking the interest group name from being visible to the seller worklet in generating an auction report. 

In this section we show that the design of the $k$-anonymity check in \fledge\ is ineffective, and the attacks can easily be adapted to maintain tracking even when the $k$-anonymity check is implemented. We show two approaches to obtain $UID$ during reporting time despite the $k$-anonymity check: controlling multiple Google accounts and employing covert channels.

\subsection{Controlling Multiple Google Accounts}
The most intuitive way to bypass the $k$-anonymity check is to create a group of fake users and associate them with the same interest group as each targeted user. 

To mitigate this type of attack, Google limits the number of \keyword{Join($\cdot$)} requests a user can perform, even if they control many browser identifiers, by requiring a one-time-use Private State Token~\cite{privatestatetoken} for each \keyword{Join($\cdot$)} request. Currently, Google only provides a Private State Token in response to requests from an authenticated Google user, and limits the number of tokens each user may obtain. Google employs a Privacy Pass protocol \cite{davidson2018privacy} to issue and redeem these tokens to prevent them from being tied back to Google accounts.

While it is relatively easy to create Google accounts and control multiple browser identifiers, this attack has limited scalability, as the number of \keyword{Join($\cdot$)} requests is limited to the number of Google accounts controlled by the adversary times the number of Private State Tokens Google issues to each account for each time period. Assume Google issues $t$ Private State Tokens to each account, in order to track $u$ users, the adversary would need to control at least $\frac{u}{t} \cdot k$ Google accounts, assuming they all get a unique browser identifier.

\subsection{Employing Covert Channels}

\begin{figure*}[tbh]
    \centering
    \includegraphics[width=0.95\textwidth]
    {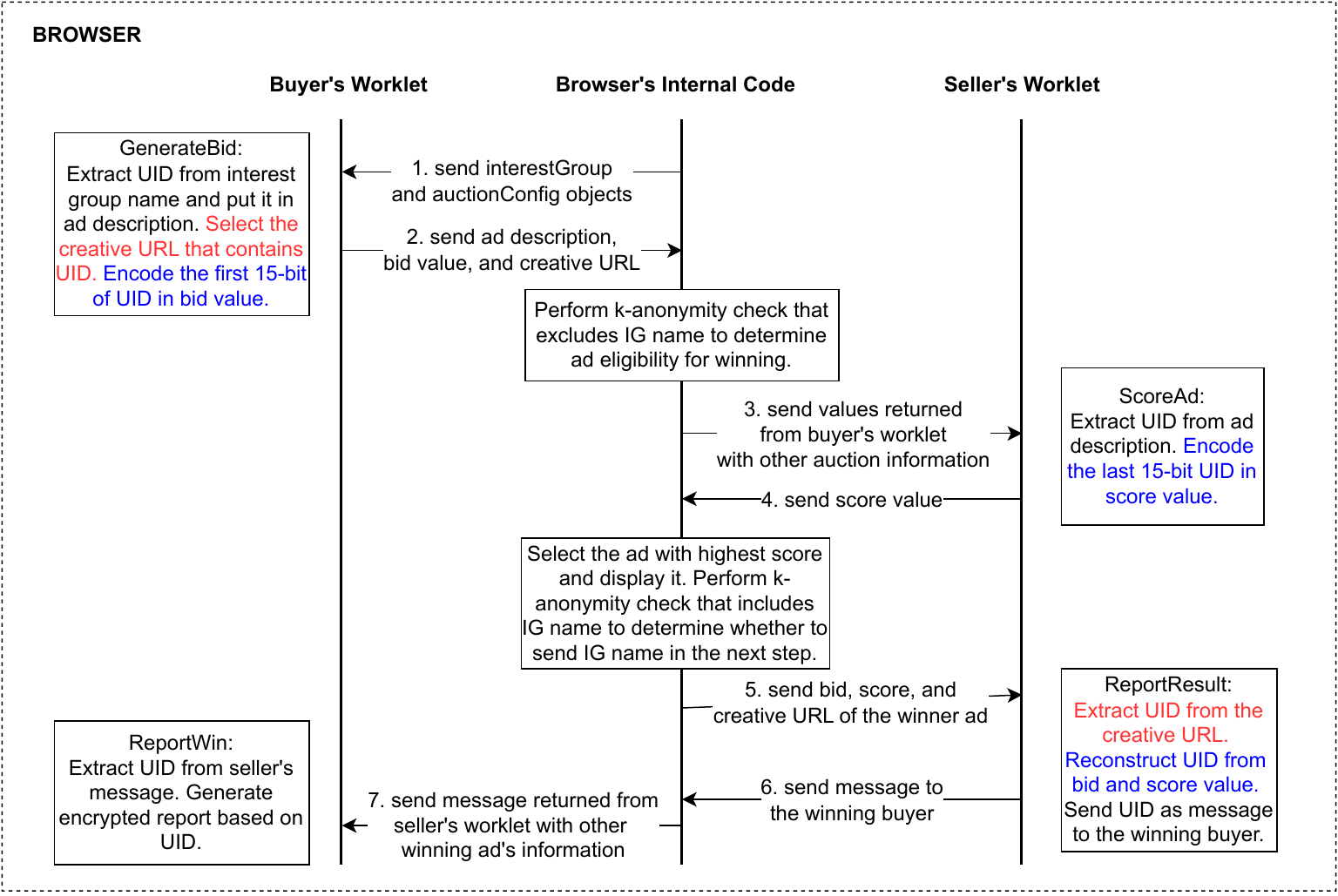}
    \caption{Using covert channels to reconstruct $UID$ during in-browser auctions. \rm Red text represents the way to extract $UID$ in creative URL. Blue text represents the way to reconstruct $UID$ through bid value and score value.}
    \label{fig:auction-process}
\end{figure*}

The more scalable way to circumvent the $k$-anonymity check is to use covert channels to re-identify the individual $UID$ without accessing interest group name during the report generation phase since the $UID$-specific interest group name will be hidden by the browser. Fortunately (for the attacker, that is), the \fledge\ protocol offers several potential covert channels to use for this.  \autoref{fig:auction-process} shows the auction process within the browser and illustrates two potential covert channels that could be used by the buyer to convey the $UID$ to the seller during the auction, as the buyer has full access to the interest group object when generating a bid. By using covert channels, the seller can extract the $UID$ in the ad creative URL or reconstruct the $UID$ through bid and score values after the failed $k$-anonymity check hides the $UID$ in the interest group name during reporting time (Step 5, \autoref{fig:auction-process}). The attributes with arrows are inputs and outputs of pre-defined worklet functions. We discuss two available covert channels below---each is sufficient by itself to reconstruct the full $UID$, but they could also be combined if the amount of information available through each channel were to be limited by mitigations.

\shortsection{Ad creative URLs} 
The $k$-anonymity check is designed to prevent a unique creative URL from winning an auction since the winning creative URL will be available to the seller during report generation phase. However, a single interest group may contain multiple ads and the buyer's code chooses which ad to bid on during the auction. This means an adversary could strategically distribute an ad tailored for one user among the ad inventory of another user's interest group, thereby undermining the $k$-anonymity threshold. Assuming an interest group can hold $A$ distinct creative URLs (\fledge\ does not appear to place any limit on $A$; we have tested up to $A = 200$), the adversary could segment the total number of tracked users $u$ by $A$, allocating a set of user-specific ads to each segment of users within their interest group. Thus, each user's interest group has a distinct name that includes the $UID$ and a set of $A$ ads, and these ads embed unique IDs within their URLs, one of which matches the $UID$. Given that $A \geq k$, the $k$-anonymity check at auction phase will be successful. When generating the bid, the buyer can choose to bid exclusively on the ad whose $UID$ matches that of the interest group name, and the colluding seller will let this ad win the auction later. In this way, the seller can directly obtain the $UID$ in the winning ad creative URL during reporting time and send it to the buyer (Step 5 and 6, \autoref{fig:auction-process}). 

\shortsection{Bid and score values}
As shown in \autoref{fig:auction-process} Step 5, the winning bid and score values are revealed to the seller for use in generating reports. Since the buyer and seller are colluding, the buyer can generate any positive floating-point number as the bid value, embed information about the $UID$ in these bits, and still secure the auction win. The seller can also encode information within the floating-point score value, as long as the score assigned to the colluding buyer's bid remains the highest among all bids, ensuring the buyer's victory in the auction. To avoid these numbers exfiltrating information from the interest group, \fledge plans to perform stochastic rounding in the browser to limit the precision to an 8-bit exponent and 8-bit mantissa. However, this still leaves room for a 32-bit information channel, 16 bits in each the bid and score, between the buyer's code and the seller's code. Suppose each user is assigned a 30-bit $UID$ in the interest group name. When generating the bid, the buyer converts the first 15 bits of $UID$ to floating point number as a positive bid value to fit within the precision available, puts the whole 30-bit $UID$ in the ad description field, and returns them along with other fields to the browser (Step 1 and 2, \autoref{fig:auction-process}). When the seller's code receives the bid, it extracts the $UID$ from ad description, converts the latter 15-bits of the $UID$ to a floating point number encoding, and returns this positive value as score to the browser (Step 3 and 4, \autoref{fig:auction-process}). For all the bids from other buyers, the seller simply returns a negative score so that the score that encodes the $UID$ is guaranteed to be the highest score and win the auction. Then, during reporting time (Step 5 and 6, \autoref{fig:auction-process}), the seller can reconstruct the $UID$ by converting the winning bid and score back to two 15-bit representations and concatenate them together to form the 30-bit $UID$, sufficient for uniquely tracking up to $2^{30}$ (over 1 billion) distinct users. 
\section{Countermeasures} \label{sec:countermeasures}

Based on our analysis, there are two types of potential countermeasures to mitigate the linking attacks. We note, however, that given the myriad opportunities available for adversaries to obtain information through the basic in-browser auction and reporting functionalities, it is difficult to have high confidence that any set of countermeasures, short of reconsidering the design and drastically limiting available communication channels, would be sufficient to eliminate tracking opportunities.

\shortsection{Limit arbitrary code execution within worklets}
To re-identify a user with the $k$-anonymity check in place, the adversary uses worklet functions to execute arbitrary code during auctions and exploits covert channels to exchange information between buyers and sellers via function outputs. A potential mitigation strategy would impose stricter controls on the format and content of these outputs. Such limits would not eliminate the cover channel, but could limit its bandwidth enough to disrupt the attack. For example, rather than allowing buyers to transmit any ad metadata to sellers during auctions, browsers could restrict communications to a selection of predefined categorical attributes. Similarly, instead of permitting sellers to assign arbitrary scores to ads, browsers could enable them to evaluate bids and ads using a fixed set of standardized rankings. Such measures would complicate adversaries' efforts to conduct widespread surveillance by limiting the possible combinations of transmitted values. However, the impact of such restrictions on advertisers is unclear, particularly regarding their ability to design bidding strategies that take advantage of available information.

\shortsection{Alternative aggregate reporting mechanisms}
In the proposed attack, the adversary takes advantage of the expectation linearity within noise addition mechanisms used by the Private Aggregation Service. To counteract this, \fledge could implement different methods for adding noise that provide increased resistance to such exploits. One potential approach is to utilize local differential privacy (LDP) \cite{dwork2008differential}, which introduces noise at the individual data point level before the aggregation process. As each piece of data is independently obscured, it will be more difficult for an adversary to neutralize the noise by aggregating the data. Local differential privacy ensures that the added noise remains effective in preserving privacy, even when data is combined. The amount of information available to an adversary could also be limited by substantially reducing the number of available bucket keys (currently $2^{128}$) and the maximum value sensitivity (which effectively reduces $\epsilon$ if it is still set based on the original value). Any limits on the information available in a report, though, reduce the potential value of reports to advertisers. % and publishers.

\section{Discussion} \label{sec:discussion}

Beyond assessing the privacy attributes in the design of \fledge, it is essential to contextualize its role within the broader advertising ecosystem. Our discussion extends to how \fledge influences competitive dynamics and the role of trusted third parties within the Privacy Sandbox framework. 

\shortsection{Market Dynamics}
While we focus on privacy aspects of \fledge, there is also a concern among competing ad networks that the design of the protocol could diminish their ability to compete effectively. They fear losing access to valuable information that is crucial for targeting and measuring ads, which might consolidate Google's dominance in the online advertising space. The UK's Competition and Markets Authority (CMA) has taken an active role in scrutinizing the Privacy Sandbox proposals, launching a formal investigation to understand their implications for competition and consumer welfare \cite{cmainvestigation}. This process involves collaboration between the CMA and the Information Commissioner’s Office (ICO), blending expertise from privacy and competition regulatory perspectives. Google has responded by offering commitments designed to ensure that the implementation of the Privacy Sandbox proposals is transparent, equitable, and does not confer an unfair data advantage to Google's own advertising services \cite{googlecommitments}. These commitments include engaging in open dialogue with the CMA and the industry, ensuring no preferential treatment for Google's products, and not using alternative identifiers or Chrome browsing histories for ad targeting. This situation underscores the complex interplay between privacy and competitive dynamics in digital markets. The CMA's public consultation process on these commitments represents a critical step in addressing these concerns, with the potential to set legally binding conditions for the design and operation of \fledge\ and other components of the Privacy Sandbox.

\shortsection{Trusted Third Parties}
In \fledge, several servers needs to be trusted to prevent information leakage during the auction. To limit exposure through these trusted servers, the service must run in a TEE on an approved cloud platform and the code implementing the trusted server must be \emph{approved} before deployment. For example, these services must not perform event-level logging or log data that would potentially identify users such as IP addresses and timestamps, and the auditor must ensure that the set of keys and the way they are updated cannot be used for user tracking or profiling purposes. At least initially, it seems that the only candidate for this auditor would be Google, although none of the available Privacy Sandbox documents acknowledge this. 

In addition, the full implementation of $k$-anonymity server requires a trusted third-party company that is not affiliated with Google to operate a relay server and will not exchange request data with them. The need for an additional trusted party poses inherent privacy risks, since this server is collecting IP addresses from the browsers submitting these requests, and can also manipulate which messages it relays and introduce other timing channels. It is unclear from the \fledge documentation what business model would be used to support the independent relay operators in the protocol. Future research work might examine the feasibility of this type of business models and measure their effectiveness. 

Finally, the code running in the web browser that manages the ad auction, including setting up the worklets for the buyer and seller code, implementing the $k$-anonymity checks, and controlling all the information passing between the worklets and external servers, is also a critical trusted component of the system. Although it is possible that the protocol will be standardize to the point where it can be implemented in independent browsers, current implementation are only in Google's Chrome browser.

\shortsection{Responsible Disclosure} 
Since our results concern privacy vulnerabilities in a proposed system that is not yet deployed, there is no immediate disclosure risk---the vulnerabilities we discuss are only risks in a future world where third-party cookies are blocked so the current tracking they enable is no longer possible. The goal of \fledge\ is to support targeted advertising in a world where third-party cookies are no longer available and the essential privacy property \fledge\ intended to provide is unlinkability of requests, and the focus of our work is analyzing how well the proposed design meets that privacy goal in a future where third-party cookies are no longer supported. Nevertheless, we did share a version of this paper with the team at Google developing \fledge. They expressed appreciation for our work, %and did not raise any technical objections to our description or conclusions, 
but have not yet shared with us any plans to change the \fledge\ design to address the issues we have identified.

\section{Conclusion} \label{conclusion}\label{sec:conclusion}

%The issue of user privacy within the advertising ecosystem is widely debated and discussed. Users are increasingly concerned about the collection and use of their personal data for targeted advertising, while advertisers increasingly rely on this data to deliver relevant and effective ads.

Google's Privacy Sandbox initiative aims to provide a balanced solution to address privacy concerns associated with web tracking while supporting the business of targeted advertising. The core idea behind the Privacy Sandbox is to create a set of APIs that enable targeted advertising through in-browser auctions. 
%without third-party cookies. \fledge\ allows advertisers to run auctions inside the client browser directly, instead of forwarding requests to external ad servers. 

Overall, the Privacy Sandbox represents a step forward in balancing privacy concerns with the needs of advertisers. Google deserves credit for their relatively transparent process in developing this initiative, which has involved seeking input from a variety of stakeholders and experts.
However, there is still a great deal of uncertainty surrounding the Privacy Sandbox. While it is already widely deployed to the general public, essential privacy mechanisms are not in place. For instance, the auditing process of Trusted Servers has not been determined and event-level clear-text reporting will be supported until at least 2026, leaving concerns about how the system will be implemented in its final design and whether it will effectively protect user privacy as intended. As such, it is crucial to provide an early analysis of the proposed system. By identifying potential risks and areas for improvement before widespread deployment, we hope to ensure that the privacy properties of the final version of the Privacy Sandbox are clearly defined and its implementation achieves them as well as possible.

%In conclusion, our work provides an overview of \fledge and analyzes potential privacy risks based on the First Origin Trial and available documentation of its intended design. It is important to approach the Privacy Sandbox with a critical eye and continue to engage in ongoing discussion and evaluation of its potential benefits and drawbacks.
\section*{Availability} \label{sec:code}
Open-source code for producing our simulation results and corresponding graphs (\autoref{fig:accuracy}, \autoref{tab:epsilon-top-k}, and \autoref{fig:fpr}) is available as \url{https://github.com/Elena6918/PrAu-Simulation}.

\section*{Acknowledgements}
\noindent
This work was partially supported by a grants  from the National Science Foundation (\#1804603 and \#2229876). The views and conclusions
expressed here are solely those of the authors and do not necessarily represent those of any other organization or institution.

\bibliographystyle{plain}
\bibliography{references}
% \input{00_main.bbl}

\appendix
\clearpage
\section{The API Calling Sequence in \FOT} \label{FOT-calling-sequence}

\begin{figure}[ht]
    \centering
    \includegraphics[width=0.99\columnwidth]
    {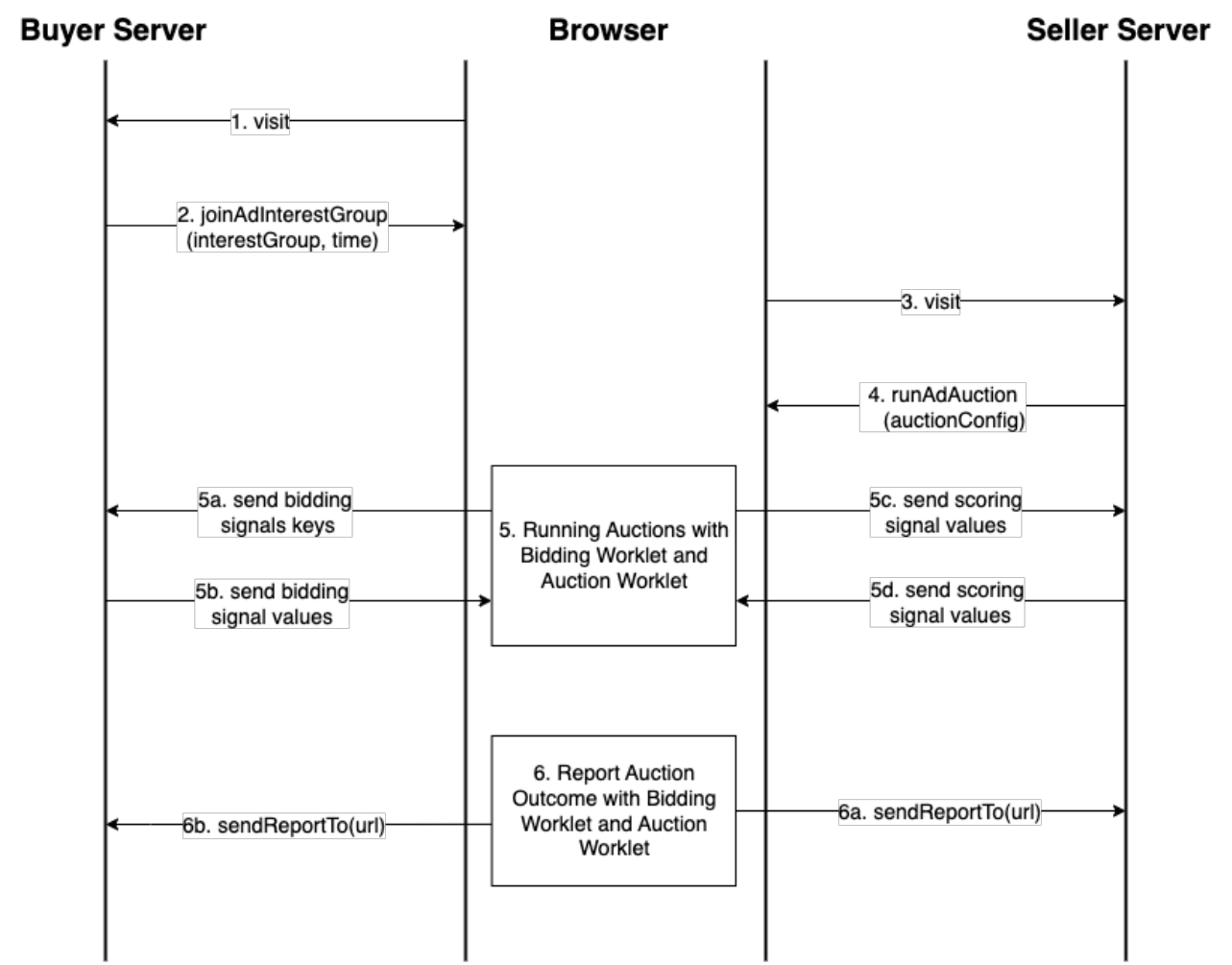}
    \caption{\FOT of \fledge with API Calling Sequence among Servers and the Browser. \rm For simplicity, we show a single seller and buyer (who is also the winner of the auction), although there would typically be many buyers and sellers.}
    \label{fig:FOT}
\end{figure}

\autoref{fig:FOT} shows the current design of \fledge in the \FOT. 

\section{Proof of Theorem 4.1} \label{theorem-explanation}

As all colluding buyers reports full $\ell_1$ contribution budget when a target user visits the seller site, we can simplify the analysis by dividing $\ell_1$ in all terms. 

The adversary implements a simple prediction algorithm: select the prediction $\hat{j}$ as the index of the value in $\vec{x}$ with the highest value. Then, the expected accuracy of this algorithm can be defined as a function $E(\epsilon, u, n)$, which is the probability $P(x_j > x_i, \forall i \neq j)$. This can be further written as $\prod_{i \in [u]-j} P(Y_i < n + Y_j)$.

Recall the probability density function (PDF) of $Y$ is $f_{Laplace}(Y) = \frac{1}{2}e^{-\epsilon|Y|}$, and the cumulative density function (CDF) is 
 \[
F_{Laplace}(Y) = \begin{cases} \frac{1}{2} e^{\epsilon Y},& \text{if } Y < 0 \\ 1- \frac{1}{2} e^{-\epsilon Y}& \text{if } Y \geq 0 \end{cases}
\]

Since each $Y_i$ is drawn independently from $\mathit{Laplace}(0, \frac{1}{\epsilon})$, we can represent $\prod_{i \in [u]-j} P(Y_i < n + Y_j)$ as \[
\int_{Y_j}F_{Laplace}(n+Y_j)^{u-1} dY_j = \int_{-\infty}^{\infty}f_{Laplace}(Y_j) \cdot F_{Laplace}(n+Y_j)^{u-1}dY_j
\]

Since $F_{Laplace}(n+Y_j)$ has two different forms depending on the sign of $Y_j+n$ and $f_{Laplace}(Y_j)$, we need to split it into two cases for calculation:

\textbf{Case 1:} $n+Y_j \geq 0$

To solve the following two integrals, we use the midpoint rule method to approximate the value. This divides the interval \([a, b]\) into \(n\) subintervals of equal width, denoted by \(\Delta x\). The midpoints of these subintervals are denoted as \(x_i^*\), where \(i = 1, 2, \ldots, n\). The approximation of the integral using the Midpoint Rule is given by:
\[ \int_a^b f(x) \, dx \approx \sum_{i=1}^n f(x_i^*) Y x \]
where \(\Delta x = \frac{b - a}{n}\).

Case 1A. $Y_j > 0$. To apply the midpoint rule to approximate $\int_{0}^{\infty} \frac{\epsilon}{2}e^{-\epsilon Y_j} \cdot (1-\frac{1}{2}e^{-\epsilon(n+Y_j)})^{u-1} dY_j$, we first discretize the integral by dividing the range $[0, \infty)$ into $L$ subintervals of equal width. Let $\Delta x$ be the width of each subinterval: $\Delta x = \frac{\infty - 0}{L}$, and the midpoints of these subintervals are denoted as $x_i = \frac{(2i-1) \Delta x}{2}$, where $i = 1, 2, \ldots, L$. For simplicity, let $\Delta x = 1$, then for each subinterval, we have $f(x_i) = \frac{\epsilon}{2}e^{-\epsilon x_i}(1-\frac{1}{2}e^{-\epsilon(n+x_i)})^{u-1}$. 

Lastly, we sum up the approximations for all subintervals to obtain an approximation for the entire integral:
% \[ \int_{0}^{\infty} \frac{\epsilon}{2}e^{-\epsilon Y_j} \cdot \left(1-\frac{1}{2}e^{-\epsilon(n+Y_j)}\right)^{k-1} dY_j \approx \sum_{i=1}^{\infty} \frac{\epsilon}{2}e^{-\epsilon(i-\frac{1}{2})} \cdot (1-\frac{1}{2}e^{-\epsilon(n+i-\frac{1}{2}))})^{k-1} \]
\begin{align}
    1A &=\int_{0}^{\infty} \frac{\epsilon}{2}e^{-\epsilon Y_j} \cdot \left(1-\frac{1}{2}e^{-\epsilon(n+Y_j)}\right)^{u-1} dY_j \\
    &\approx \sum_{i=1}^{\infty} \frac{\epsilon}{2}e^{-\epsilon(i-\frac{1}{2})} \cdot (1-\frac{1}{2}e^{-\epsilon(n+i-\frac{1}{2}))})^{u-1}
\end{align}
Case 1B. $Y_j < 0$. To apply the Midpoint Rule to approximate $\int_{-n}^{0} \frac{\epsilon}{2}e^{-\epsilon|Y_j|} \cdot (1-\frac{1}{2}e^{-\epsilon(n+Y_j)})^{u-1} dY_j$, we first discretize the integral by dividing the range $[-n, 0]$ into $n$ subintervals of equal width $\Delta x = 1$. Then, the midpoints of these subintervals can be expressed as $x_i = -n + \frac{(2i-1)}{2}$, where $i = 1, 2, \ldots, n$. Thus, for each subinterval, we have $f(x_i) = \frac{\epsilon}{2}e^{-\epsilon x_i}(1-\frac{1}{2}e^{-\epsilon(n+x_i)})^{u-1}$.

Lastly, we sum up the approximations for all subintervals to obtain an approximation for the entire integral:
% \[ \int_{-n}^{0} \frac{\epsilon}{2}e^{-\epsilon|Y_j|} \cdot (1-\frac{1}{2}e^{-\epsilon(n+Y_j)})^{u-1} dY_j \approx \sum_{i=1}^n \frac{\epsilon}{2}e^{\epsilon(-n+i-\frac{1}{2}))} \cdot (1-\frac{1}{2}e^{-\epsilon(i-\frac{1}{2}))})^{u-1} \]
\begin{align}
    1B &= \int_{-n}^{0} \frac{\epsilon}{2}e^{-\epsilon|Y_j|} \cdot (1-\frac{1}{2}e^{-\epsilon(n+Y_j)})^{u-1} dY_j \\
    &\approx \sum_{i=1}^n \frac{\epsilon}{2}e^{\epsilon(-n+i-\frac{1}{2}))} \cdot (1-\frac{1}{2}e^{-\epsilon(i-\frac{1}{2}))})^{u-1}
\end{align}

\textbf{Case 2:} $n+Y_j<0$

Case 2A. $Y_j > 0$. Given $n > 0$ and $Y_j > 0$, it is impossible that $Y_j > 0$, and thus Case 2A is invalid.

Case 2B. $Y_j < 0$

\begin{align}
2B &= \int_{-\infty}^{-n} \frac{\epsilon}{2}e^{-\epsilon|Y_j|} \cdot (\frac{1}{2}e^{\epsilon(n+Y_j)})^{u-1} \, dY_j \\
&= \int_{-\infty}^{-n} \frac{\epsilon}{2}e^{\epsilon Y_j} \cdot \frac{1}{2^{u-1}}e^{\epsilon(u-1)(n+Y_j)}dY_j \\
&= \frac{\epsilon}{2^u} \cdot [\frac{1}{\epsilon u}e^{\epsilon uY_j+\epsilon(u-1)n}]_{-\infty}^{-n} \\
&= \frac{1}{2^u \cdot u}e^{-\epsilon n}
\end{align}

Lastly, after summing up all 4 cases, we can evaluate the integral as the following form:
\begin{align}
    Accuracy = &\sum_{i=1}^{\infty} \frac{\epsilon}{2}e^{-\epsilon(i-\frac{1}{2})} \cdot (1-\frac{1}{2}e^{-\epsilon(n+i-\frac{1}{2}))})^{u-1} \; &\text{Case 1A} \\
    &+ \sum_{i=1}^n \frac{\epsilon}{2}e^{\epsilon(-n+i-\frac{1}{2}))} \cdot (1-\frac{1}{2}e^{-\epsilon(i-\frac{1}{2}))})^{u-1} \; &\text{Case 1B} \\
    &+ \frac{1}{\epsilon \cdot u \cdot 2^u} e^{-\epsilon n} \; &\text{Case 2B}
\end{align}

\end{document}